\documentclass[twocolumn,showpacs,preprintnumbers,amsmath,amssymb,aps]{revtex4-2}
\usepackage{graphicx}
\usepackage{dcolumn}
\usepackage{bm}
\usepackage[colorlinks=true]{hyperref}
\usepackage{epstopdf}
\usepackage[T1]{fontenc} 
\usepackage{lmodern}
\usepackage{braket}
\usepackage{color}

\begin{document}

\title{Interface disorder as a cause for kinetic Rashba-Edelstein effect and interface Spin-Hall effect at the metal - insulator boundary.  }

\author{A.~V.~Shumilin, V.~V.~Kabanov}

 \affiliation{ Jozef Stefan Institute, 1000 Ljubljana, Slovenia }

\begin{abstract}
The spin phenomena observed at a clean metall-insulator interface are typically reduced to Rashba-Edelstein effect, that leads to spin accumulation over a few monolayers.
We demonstrate that the presence of interface disorder significantly expands the range of potential phenomena.  Specifically, the skew scattering at the metal - insulator boundary
gives rise to the "kinetic Rashba-Edelstein effect", where spin accumulation occurs on a much larger length scale comparable to mean free path.
Moreover, at higher orders of spin-orbit interaction, skew scattering is accompanied with spin relaxation resulting in the interface spin-Hall effect -
a conversion of electrical current to spin current at the metal surface. Unlike the conventional spin-Hall effect, this phenomenon persists even within the Born approximation.
These two predicted phenomena can dominate the spin density and spin current in devices of intermediate thickness.
\end{abstract}

\maketitle
\section{Introduction}
\label{sec:intro}
Spin-orbit interaction is a fundamental phenomenon that enables the exchange of angular momentum between orbital and spin degrees of freedom.
One of the promising applications of this interaction is in the development of spin-orbit torque devices, which offer a scalable and field-free solution for magnetic memory that can be controlled by electrical currents \cite{RecAdv,BigRev, Comm1}.
Experimental studies have already demonstrated the magnetization switching  \cite{SwMiron2011,SwLiu2012,SwOh2016} and domain wall motion \cite{WallMiron2011} in such devices, with other potential applications including the control of magnetic skyrmions \cite{Skyrmion} and spin-waves \cite{Chumak2015}.

The basic setup for an spin-orbit torque device involves a magnetic bilayer composed of a heavy metal layer with strong spin-orbit coupling and a ferromagnetic layer that acts as the detector for spin polarization and spin current from the heavy metal. Although this simple bilayer composition is already functional \cite{YIGPt}, additional layers are often added to enhance or modify the device properties.  These additional layers can be added from the side of the ferromagnet \cite{oxides}, from the side of heavy metal \cite{PyPtCo}, or between them \cite{spacer}. They are usually composed from materials possessing specific spin properties, such as antiferromagnets \cite{AFM} and topological insulators  \cite{BiSeAg, BiSeTa}.

Surprisingly, recent research has shown that even insulating non-magnetic molecules can significantly enhance the spin torque when added from the side of a heavy metal, for the heavy metal thicknesses up to 5nm \cite{molecules}. These results have drawn our attention to the spin phenomena occurring at the interface between heavy metals and insulators.

\begin{figure}[t]
  \centering
  \includegraphics[width=0.4\textwidth]{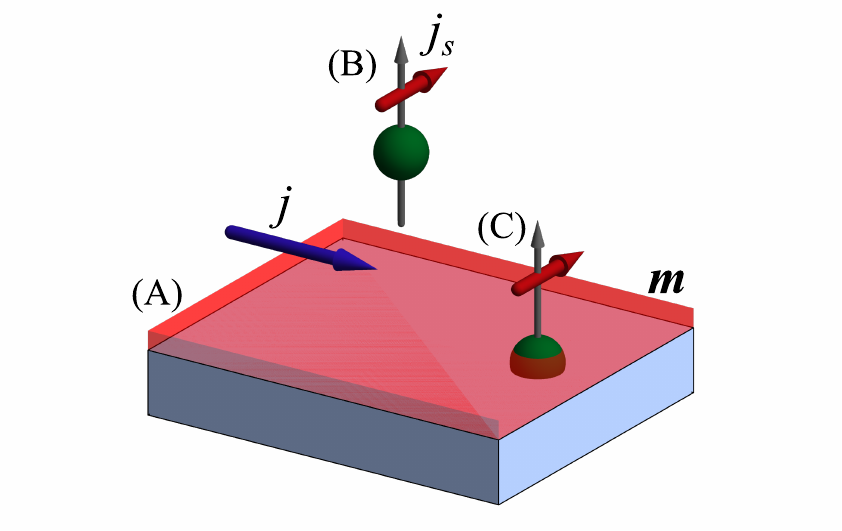}
  \caption{The three phenomenological effects leading to the spin polarization in the heavy metal layer. (A) Rashba-Edelstein effect, (B) Spin-Hall effect, (C) Interface spin-Hall effect.
  The blue arrow stands for the direction of electrical current. Grey arrows show the direction of spin flow. The red arrows correspond to the direction spin polarization. Green spheres depict
  the impurities.
  }
  \label{fig:pic}
\end{figure}

Spin generation in magnetic bilayers is typically attributed to either the Rashba-Edelstein surface effect or the Spin-Hall effect (see Fig.~(\ref{fig:pic})).
Although it can be difficult to distinguish between these effects in a given experiment \cite{SwMiron2011}, they differ significantly in terms of the device engineering.
The Rashba-Edelstein effect results in spin polarization at the surface of heavy metal where inversion symmetry is broken. However, this polarization is confined to only a few monolayers near the interface \cite{KrasChu, KrasTeorem, Krasovskii2015, DrogTok}, meaning that the spin generated at the heavy metal interface with a third material can affect the ferromagnetic layer only in very thin devices.
When the heavy metal layer is thick, the spin-torque in the ferromagnetic layer is usually attributed to Spin-Hall effect, which is the conversion of electrical to spin current in the heavy metal layer \cite{PtThick, PtSH}.
However, the Spin-Hall effect is related to the bulk properties of the heavy metal \cite{RashbaRev,DiakBook,SinovaRev} and is not expected to be significantly influenced by the heavy metal - insulator interface. In relatively clean samples Spin-Hall effect is dominated by skew scattering at impurities in the bulk, that is absent in the Born approximation, resulting in a suppression factor of $V_0/\varepsilon_F$ where $V_0$ is the potential of single impurity and $\varepsilon_F$ is Fermi energy.

The recently predicted interface Spin-Hall effect combines the properties of Rashba-Edelstein and conventional Spin-Hall effects. It refers to the electrical to spin current conversion at the interface. Its phenomenological possibility is demonstrated in \cite{AminFor, AminPhen, TokatGran}. However, it has only been studied for the interface of two metals and was attributed to spin filtering, which involves different probabilities for spin-up and spin-down electrons to traverse the interface between metals \cite{Baek2018, AminDFT,BigRev}. Similar spin-filtering was also predicted for tunneling through semiconductor barriers \cite{Vosk1988, Tar2003, Gla2005}.

Here we investigate the spin kinetics near a disordered heavy metal-insulator interface. We demonstrate that the disorder significantly increases the variety of interface spin phenomena. Skew scattering at the interface impurities causes spin accumulation over a distance comparable to the mean free path from the interface. This phenomenologically corresponds to the Rashba-Edelstein effect; however, the thickness of the spin accumulation layer is significantly larger than that predicted in \cite{KrasTeorem, Krasovskii2015}. Combined with spin relaxation, this leads to the interface Spin-Hall effect, which is absent at a clean metal-insulator interface. Both of these phenomena are sensitive to the materials that make up the interface and their properties, as well as their disorder.

\section{Model of the heavy metal surface}
\label{sec:bord}

We consider a clean heavy metal interface with insulator described by the model Hamiltonian
\begin{equation}\label{Ham}
 \widehat{H} = \frac{\widehat{\bf p}^2}{2m}  + U(z) - \gamma \frac{\partial U}{\partial z}\left(\sigma_x \widehat{p}_y - \sigma_y \widehat{p}_x \right)
\end{equation}
Here $\widehat{\bf p} = -i\hbar\nabla$ is the momentum operator, $m$ is the effective mass, $U(z) = U_0\theta(z)$ is the potential energy describing the abrupt barrier at $z=0$ with the height $U_0$. $\gamma$ is  effective spin-orbit interaction inside heavy metal. It's solution in Appendix~\ref{App1} leads to the following electron wavefunction on the metal side of the interface
\begin{equation}\label{psi1}
\Psi_{\alpha}({\bf k}) = \frac{e^{i {\bf k_\perp}{\bf r}_\perp}}{\sqrt{2V}}
\left(e^{i|k_z|z} + \hat{r}({\bf k}) e^{-i|k|_zz} \right)u_\alpha.
\end{equation}
Here ${\bf k}$ is electron wavevector, $k_z$ is its component along $z$, ${\bf k_\perp}$ is its $xy$-component and $u_\alpha$ is arbitrary spinor.
Eq.~(\ref{psi1}) includes the spin-dependent reflection amplitude $\hat{r}({\bf k})$
\begin{equation} \label{r1}
\hat{r}({\bf k}) = - e^{2i\phi_0} \left[\cos\left(\Delta \phi\right) \hat{1} + i\sin\left(\Delta \phi\right) \hat{\sigma}_{\bf k}  \right],
\end{equation}
where $\phi_0 = (\phi_+ + \phi_-)/2$ is the average phase change during the reflection and $\Delta\phi = \phi_+ - \phi_-$ shows its spin dependence.
$\phi_{\pm} = \arctan[k_z/(\kappa \mp  2m\gamma U_0 k_\perp/\hbar)]$, $\kappa = \sqrt{2mU_0/\hbar^2 - k_z^2}$.
 $\hat{\sigma}_{\bf k}$ is the combination of Pauli matrices.
\begin{equation}
\hat{\sigma}_{\bf k} = \frac{k_y}{|k_\perp|} \hat{\sigma}_x - \frac{k_x}{|k_\perp|} \hat{\sigma}_y
\end{equation}

It is demonstrated in Appendix~\ref{App1} that in agreement with previous studies \cite{KrasTeorem, Krasovskii2015} the clean interface does not exhibit an interface spin Hall effect and the spin polarization is confined to several monolayers.

The primary objective of this study is to incorporate interface disorder into the theory. The most common approach to
the electron kinetics near disordered interface involves finite probabilities of specular and non-specular reflection \cite{fuchs}.
 However, to address the new spin phenomena it is crucial
to consider a microscopic mechanism responsible for non-specular reflection. Two conventional approaches exist for modeling reflection from disordered interfaces: roughness of the interface \cite{falk,vasko} and interface impurities \cite{greene,baskin,Glazov2022}. Both approaches allow for the possibility of non-specular reflection, and we consider them to be interchangeable. In this work we adopt the latter one.

A single impurity can be described with potential energy $V_0({\bf r})$. With respect to the spin-orbit interaction it leads to the additional term in the  electron Hamiltonian
\begin{equation}\label{Vimp}
V({\bf r}) = V_0({\bf r}) + \gamma \bm{\sigma} \left[ \frac{\partial V_0}{\partial {\bf r}} \times {\bf p} \right]
\end{equation}
In this work we consider small impurities with $V_0({\bf r}) = V_I \delta({\bf r} )$ where $V_I$ stands for the magnitude of impurity potential.

\begin{figure}[t]
  \centering
  \includegraphics[width=0.45\textwidth]{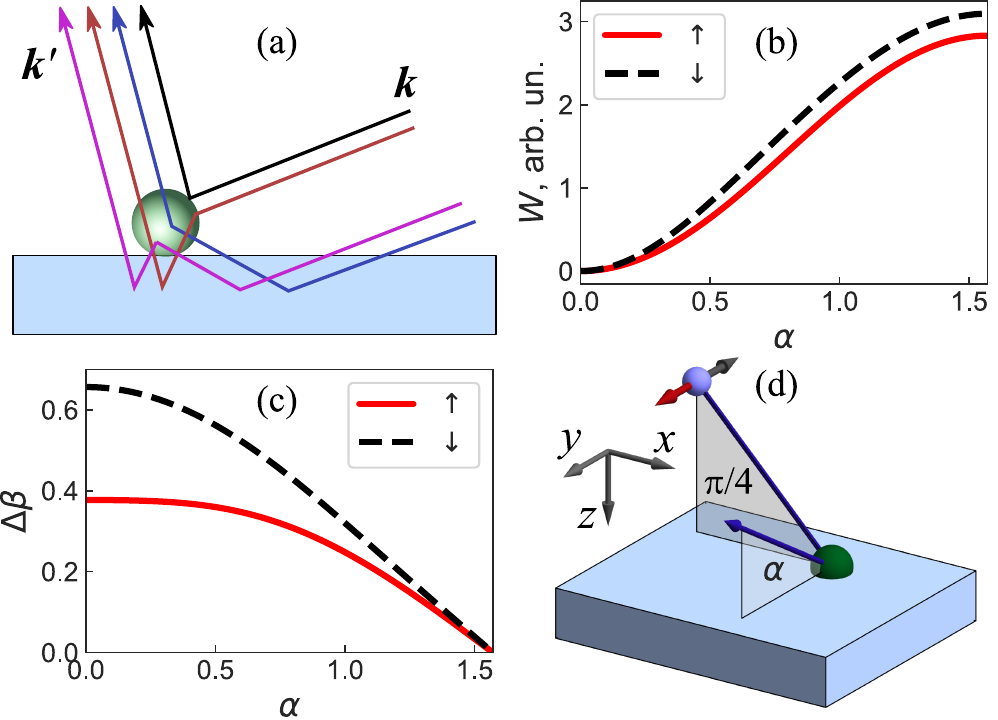}
  \caption{(a) The four possibilities for the impurity scattering from state ${\bf k}$ to the state ${\bf k}'$ in the presence of the interface. (b) the spin dependence of scattering amplitude. (c) asymmetric spin rotation angle. The results presented in panels (b) and (c) correspond to $U_0/\varepsilon_F = 4$, $\gamma p_F^2/\hbar = 0.2$ and to
  the geometry shown in panel (d). }
  \label{fig:scat}
\end{figure}

The scattering from the impurities is characterized by the matrix elements $\hat{V}({\bf k}_1,{\bf k}_2) =  V_{\alpha\beta}({\bf k}_1,{\bf k}_2) = \langle \Psi_{\alpha}({\bf k}_1) |V({\bf r})|
\Psi_{\beta}({\bf k}_2) \rangle$, where $\Psi_{\alpha}({\bf k}_1)$ and $\Psi_{\beta}({\bf k}_2)$ correspond to the electron states at the clean interface, as described by Eq.~(\ref{psi1}).
In order to analyze the scattering process, it is useful to decompose the scattering elements into the two terms.
\begin{subequations}
   \label{Vimp-all}
  \begin{equation}
    \hat{V}({\bf k}_1,{\bf k}_2) = \hat{V}^{(N)}({\bf k}_1,{\bf k}_2) + \hat{V}^{(SO)}({\bf k}_1,{\bf k}_2)
  \end{equation}
  \begin{equation} \label{Vimp0}
    V_{\alpha\beta}^{(N)}({\bf k}_1, {\bf k}_2 ) = \\ \frac{1}{2} \frac{V_I}{V} (1 + \hat{r}^+({\bf k}_1))(1+\hat{r}( {\bf k}_2))
  \end{equation}
\begin{multline}\label{VimpSO}
  \hat{V}^{(SO)}({\bf k}_1,{\bf k}_2) = \frac{i V_I \gamma}{2\hbar V} \left(
  {\bm \sigma}[{\bf p}_{1}^{(inc)} \times {\bf p}_{2}^{(inc)}]
   \right. \\
   +  \widehat{r}^+{({\bf k}_1)}{\bm \sigma}[{\bf p}_{1}^{(ref)} \times {\bf p}_{2}^{(inc)}]  +
  {\bm \sigma}[{\bf p}_{1}^{(inc)} \times {\bf p}_{2}^{(ref)}]  \widehat{r}({\bf k}_2)  \\
  \left.+
  \widehat{r}^+{({\bf k}_1)}{\bm \sigma}[{\bf p}_{1}^{(ref)} \times {\bf p}_{2}^{(ref)}] \widehat{r}({\bf k}_2) \right)
\end{multline}
\end{subequations}
Here $V_{\alpha\beta}^{(N)}({\bf k}_1,{\bf k}_2)$ corresponds to the first term in r.h.s. of Eq.~(\ref{Vimp}) and $V_{\alpha\beta}^{(SO)}({\bf k}_1,{\bf k}_2)$ is related to the spin-orbit correction and to the second term in r.h.s. of
Eq.~(\ref{Vimp}).  ${\bf p}^{(inc)} = \hbar(k_x,k_y,|k_z|)$ is the momentum of the incident electron and ${\bf p}^{(ref)} = \hbar(k_x,k_y,-|k_z|)$ is the momentum of reflected electron.
$\widehat{r}^+{({\bf k})}$ is the Hermitian conjugate of the reflection amplitude described with Eq.~(\ref{r1}).

Eqs.~(\ref{Vimp0},\ref{VimpSO})  can be represented as the four different reflection possibilities shown in Fig.~\ref{fig:scat}(a). The electron can undergo scattering from the impurity without interaction with the surface, or can be specularly reflected before or after the scattering or both.

The probability of scattering depends on the quantum interference between the different reflection possibilities, which can be constructive or destructive depending on the phase $\varphi_0 \pm \Delta\varphi$ and the relation between $\hat{V}^{(N)}$ and $ \hat{V}^{(SO)}$. This probability varies for different spin projections, leading to skew scattering and spin separation at the surface. Fig.\ref{fig:scat}(b) shows the probability for electrons with the incident angle of $\pi/4$ in the xz-plane to scatter to the yz-plane. The scattering rate depends on the spin projection to the y-axis.
The calculation details are presented in the Appendix~\ref{App2}.  When spin-orbit interaction is present, electron scattering also results in spin rotation, leading to spin relaxation after multiple random scattering events \cite{vasko}. Fig.\ref{fig:scat}(c) demonstrates that in our case, this rotation becomes asymmetric, causing the spin-up to rotate differently than the spin-down. This asymmetric spin rotation produces spin polarization of reflected electrons, even when the incident electrons are not spin-polarized, resulting in the interface spin-Hall effect. The geometry corresponding to Figs.~\ref{fig:scat}(b,c) is shown in Fig.\ref{fig:scat}(d).

\section{Spin current and spin polarization}
To understand impact of the skew scattering and asymmetric spin rotation on the electrons in the bulk of heavy metal we introduce Boltzmann equation
\begin{equation}\label{Bol1}
\frac{\partial \hat{f}({\bf r},{\bf p})}{\partial t} + {\bf v}\frac{\partial \hat{f}({\bf r},{\bf p})}{\partial {\bf r}}
+ {\bf F}\frac{\partial \hat{f} ({\bf r},{\bf p})}{\partial {\bf p}} = I(\hat{f}({\bf r},{\bf p}))
\end{equation}
Here  the distribution function $\hat{f}({\bf r},{\bf p})$ is a $2\times 2$ matrix
in spin space that depends on the coordinate and momentum as usual. $I(\hat{f}({\bf r},{\bf p}))$ is the scattering operator.

We consider the following ansatz for the distribution function: $\hat{f} = \hat{f}_0 + \hat{f}_1 + \hat{f}_2$. Here $\hat{f}_0$ is the equilibrium electron distribution. It is proportional to the unit matrix $\hat{1}$ in the spin space.
\begin{equation}\label{f1}
 \hat{f}_1  = - \frac{j}{e}\frac{p_x}{n} \frac{\partial \hat{f}_0 }{\partial \varepsilon}
\end{equation}
represents the electric current density $j$ which is assumed to flow along the $x$-axis.
$\hat{f}_2$ describes the spin polarization. We assume $\hat{f}_2 \ll \hat{f}_1\ll \hat{f}_0$,
 which corresponds to a relatively small spin polarization due to the skew scattering.
In this case it is possible to neglect $f_2$ for the incident electrons, because it would lead only to
a small correction for $f_2$ of scattered electrons (that is responsible for kinetic Rashba-Edelstein and interface spin-Hall effects).

The skew scattering should be introduced into Boltzmann equation as a boundary condition. It is derived in Appendix~\ref{App3} and reads
\begin{multline}\label{f2}
\hat{f}_2({\bf p}, z=0) = \frac{m V}{S|p_z|} \hat{r}({\bf p}) \left( \int\frac{V d{\bf p}'}{(2\pi \hbar)^3}  \times  \right.
\\
\left.
\widehat{\cal W} \left(\frac{{\bf p}}{\hbar},\frac{{\bf p}'}{\hbar}\right)
\bigl(f_1({\bf p}') - f_1({\bf p}) \bigr)  \right)  \hat{r}({\bf p})^+
\end{multline}
Here
\begin{equation}\label{calW}
\widehat{\cal W}({\bf k},{\bf k}') = \frac{2 \pi}{\hbar} N_I \widehat{V}({\bf k},{\bf k}')\widehat{V}({\bf k}',{\bf k})\delta(\varepsilon_k - \varepsilon_{k'})
\end{equation}
$N_I$ is the total number of impurities.

Eqs.~(\ref{f2},\ref{calW}) show that $f_2$ is proportional to the two-dimensional impurity concentration $N_I/S$ which controls the probability $P_{nsp}$ that an incident electron  with Fermi energy is reflected non-specularly.
This probability, averaged over incident electron momenta, can be expressed as follows:
\begin{subequations}
\begin{equation}\label{Pnsp}
P_{nsp} = \frac{4\pi^2\hbar^3}{S  p_F^2}{\cal I}_N
\end{equation}
\begin{equation}
{\cal I}_N = \int \int \frac{V^2 d{\bf p}d{\bf p}'}{(2 \pi \hbar)^6} {\rm Tr}  \widehat{\cal W}\left(\frac{\bf p}{\hbar},\frac{{\bf p}'}{\hbar}\right) \delta(\varepsilon - \varepsilon_F)
\end{equation}
\end{subequations}

In our model the macroscopic symmetry in $xy$-plane is not broken, and the only possible spin current density $j_s$ in $z$ direction
describes the flow of y-polarized spins.  It is conventionally expressed with the interface spin-Hall angle
\begin{subequations}\label{spinhall}
\begin{equation}
  \tan \theta_{sh} = \frac{j_s}{j/e} = \frac{3}{4} P_{nsc} \frac{{\cal I}_{sh}}{{\cal I}_N}
\end{equation}
\begin{multline}
{\cal I}_{sh} = \int \frac{V^2 d{\bf p}d{\bf p}'}{(2 \pi \hbar)^6} \left(\frac{p_x'}{p_F} - \frac{p_x}{p_F} \right)\delta(\varepsilon - \varepsilon_F)  \times \\
{\rm Tr} \hat{\sigma}_y\left[ \hat{r}({\bf p}) \widehat{\cal W}\left(\frac{\bf p}{\hbar},\frac{{\bf p}'}{\hbar}\right)  \hat{r}^+({\bf p})  \right]
\end{multline}
\end{subequations}

\begin{figure}[t]
  \centering
  \includegraphics[width=0.4\textwidth]{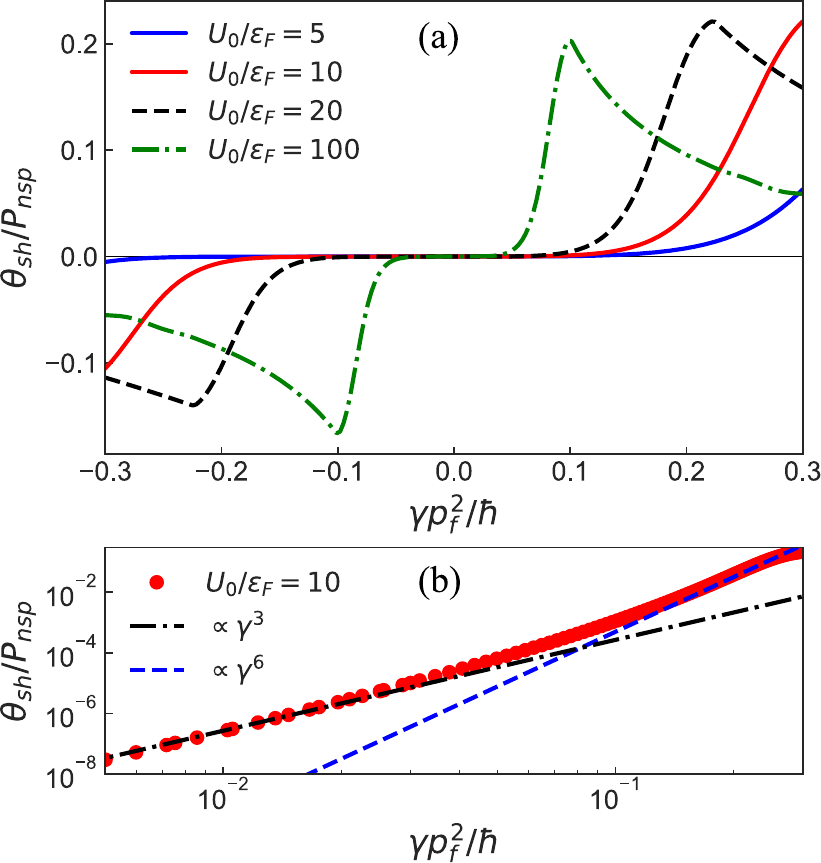}
  \caption{(a) Interface Spin-Hall angle as a function of spin-orbit interaction for various relationships between interface potential and Fermi energy $U_0/\varepsilon_F$ as indicted in the legend.
   (b) The $\theta_{sh}(\gamma)$ dependence for $U_0/\varepsilon_F=10$ in the double logarithm scale. The blue dashed line and the black dashed-dotted line correspond to $\theta_{sh}\propto \gamma^6$ and
  $\theta_{sh}\propto \gamma^3$ respectively.  }
  \label{fig:SH}
\end{figure}

The interface spin-Hall angle calculated with Eq.~(\ref{spinhall}) is shown in Fig.~\ref{fig:SH} (a)
as a function of spin-orbit interaction for different barrier heights $U_0$. Interestingly, $\theta_{sh}$ is not an odd function of $\gamma$. It is related to  the
electron momentum ${\bf p}$ being an operator that does not commute with impurity potential $V_0({\bf r})$.
Fig.~\ref{fig:SH} (b) shows $\theta_{sh}(\gamma)$ dependence in the double logarithm scale for $U_0 = 10 \varepsilon_F$ and small positive $\gamma$. The spin-Hall angle is a high-order function of $\gamma$, $\theta_{SH} \propto \gamma^3$ at very small $\gamma$ and becomes steeper with the increase of $\gamma$, up to $\theta_{SH} \propto \gamma^6$. As shown in Fig.~\ref{fig:SH} (a)
there is a maximum in this dependence corresponding to $\gamma p_F^2/\hbar \sim \sqrt{\varepsilon_F/U_0}$.

The physical mechanism of interface spin Hall effect consists of the two parts. The first one is the skew scattering shown in Fig.~\ref{fig:scat}(b). However, the skew scattering alone is insufficient to generate the spin current if spin is conserved during the scattering. Therefore, the interface spin Hall requires also the asymmetric spin rotation shown in Fig.~\ref{fig:scat}(c) that leads to the "spin relaxation" which depends on spin projection. The spin relaxation is the second order in $\gamma$ phenomenon \cite{vasko}, meaning that the interface spin-Hall effect is absent in the first order approximation over $\gamma$.

To show that the spin separation itself is a first order effect,  we examine the spin accumulation near the interface. We solve the Boltzmann equation using the minimal model, which assumes that the scattering operator in the bulk can be described by a single relaxation time $\tau$: $I(\hat{f}) = (\hat{f}_0 - \hat{f})/\tau$.
By applying this assumption to Eq.~(\ref{Bol1}), we obtain the solution
\begin{equation}\label{f2z}
\hat{f}_2({\bf p},z) = \hat{f}_2({\bf p},0) \exp\left( - \frac{z p_f }{p_z l_{free}} \right)
\end{equation}
Here $l_{free} = v_F\tau$ is the mean free path.

The spin polarization calculated from Eq.~(\ref{f2z}) reads
\begin{subequations}
\begin{equation}\label{m3}
m_y(z) = \frac{3}{4} \frac{j}{ev_F} P_{nsp} \frac{{\cal I}_{s}(z)}{{\cal I}_{N}}
\end{equation}
\begin{multline}\label{Is}
{\cal I}_{s}(z) = \int \frac{V^2 d{\bf p}d{\bf p}'}{(2\pi \hbar)^6}
\times \\
  {\rm Tr} \sigma_y   \hat{r}({\bf p})   \widehat{\cal W}\left(\frac{\bf p}{\hbar},\frac{{\bf p}'}{\hbar}\right)   \hat{r}^+({\bf p})  \times \\
 \left( \frac{p_x'}{|p_z|} - \frac{p_x}{|p_z|}\right) \exp\left( - \frac{z p_f}{l_{free}|p_z|}\right) \delta(\varepsilon- \varepsilon_F)
\end{multline}
\end{subequations}

\begin{figure}[t]
  \centering
  \includegraphics[width=0.4\textwidth]{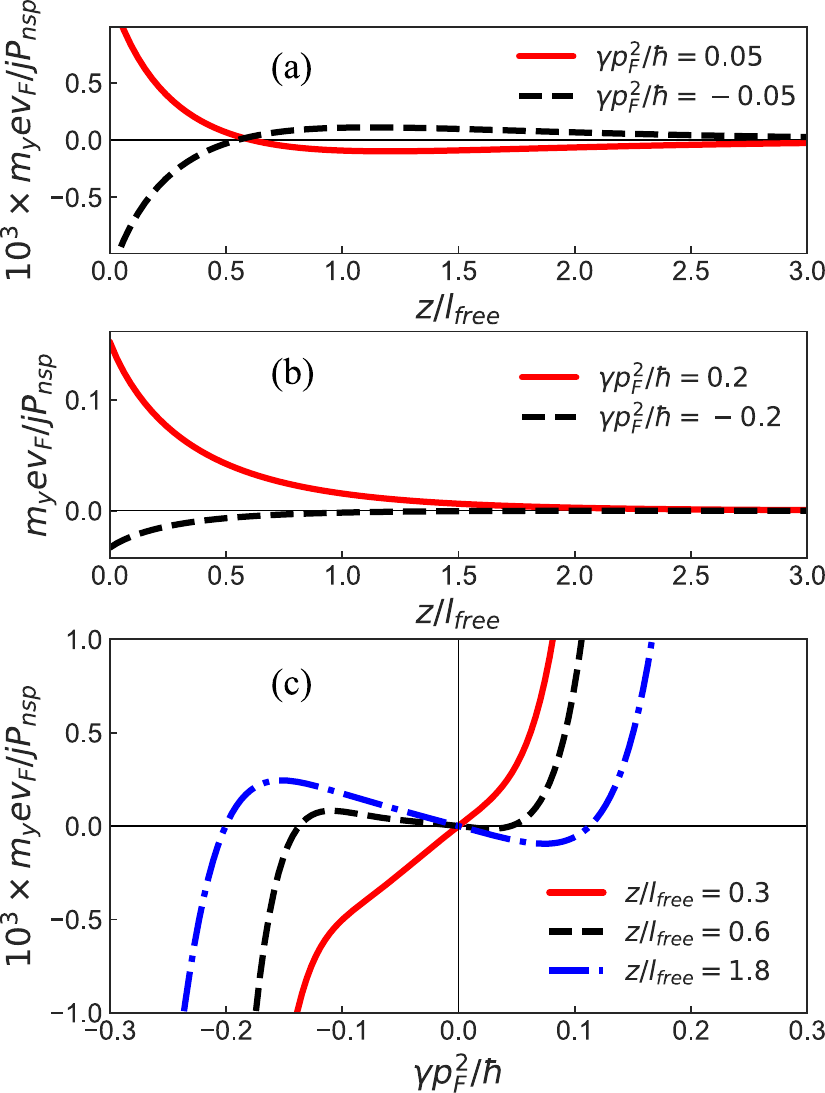}
  \caption{Spin accumulation $m_y$ due to kinetic Rashba-Edelstein effect calculated for $U_0 = 10\varepsilon_F$.  Panels (a) and (b) show the dependence $m_y(z)$ for different values of $\gamma$. Panel (c) shows the dependence of accumulated spin on the spin orbit interaction parameter $\gamma$ for three different values $z/l_{free}$. }
  \label{fig:S}
\end{figure}

Fig.~\ref{fig:S} shows the spin polarization $m_y$ calculated for $U_0 = 10\varepsilon_F$. Similarly to the interface spin-Hall effect, the polarization obtained from the Boltzmann equation
appears only due to the possibility of non-specular reflection and is normalized by $P_{nsc}$.
Fig.~\ref{fig:S}(a) and (b) show the $m_y$ distribution over $z$ for $\gamma p_F^2/\hbar = \pm 0.05$ and $\gamma p_F^2/\hbar = \pm 0.2$ correspondingly. For small values of $\gamma$, $m_y$ changes its sign as a function of $z$ indicating the difference in the average reflection angle for different spin projections. This change of the sign disappears at larger values of $\gamma$ when interface spin-Hall effect becomes sufficiently strong to dominate the spin accumulation.

Phenomenologically, the calculated spin accumulation is the Rashba-Edelstein effect because it is confined near the interface. However, unlike the conventional
Rashba-Edelstein effect at clean interfaces studied in \cite{KrasTeorem, Krasovskii2015}, the polarization spans up to the mean free path $l_{free}$. In the case where the sample only contains impurities at the surface and the bulk is clean, $l_{free}$ can be arbitrary large. Thus, the resulting spin polarization can be referred as the kinetic Rashba-Edelstein effect.

\section{discussion}

The interface spin Hall and kinetic Rashba-Edelstein effects are closely related to the interference of different reflected waves shown in Fig.~\ref{fig:scat} (a). Both the effects should disappear if this interference is somehow suppressed. For example one can consider bulk impurities near the interface as a possible reason for interface spin Hall. However, in this case the phase of the reflection amplitude $\hat{r}$ would be modified by the random distance between impurity and the surface. If this distance exceeds $\hbar/p_f$, the interference and the spin generation would be suppressed. Nevertheless, if the impurities are located precisely at the interface and are small compared to $p_F$, their potential $V_I$ is not significant: it is absorbed into non-specular reflection probability $P_{nsp}$ in our final expressions. This is a consequence of the calculations performed within Born approximation. We presume that introduction of complex impurities with large size and high potential energy can significantly modify the spin accumulation and spin current.

The existence of interface spin-Hall effect in Born approximation makes it fundamentally different from the ordinary spin-Hall effect that appears only as a higher-order corrections over impurity potential $V_I$
and is suppressed if the disorder is Gaussian. Although there are some predicted mechanisms for its existence in Gaussian disorder, namely the combined scattering from impurities and phonons \cite{valley} and the scattering at close impurity complexes \cite{valley,Dmitriev}, all these mechanisms require high orders of perturbation theory. Interface spin Hall is possible in the Born approximation because of the large interface potential energy that exists already in zeros order.

The unusual properties of the kinetic Rashba-Edelstein effect and interface spin-Hall effect suggest that in some cases the predicted interface phenomena should dominate the spin accumulation and spin current. It happens when the thickness $d$ of the sample is intermediate $\hbar/p_F \ll d \lesssim l_{free}$ and bulk impurities that control conductivity have small potential energy $V_0 \ll \varepsilon_F$.
If these conditions are met, the control of interface properties is important to optimize the spin-torque. The effect of interface roughness is sometimes reported in experiments \cite{molecules,rough}, however, to the best of our knowledge, there is no its systematic study. We predict that the perfect interface is not necessarily the clean one. The controlled disorder  can serve as a source of both spin polarization and spin current. It allows to control the functionality of spin-torque devices by surface engineering, i.e. by manipulating surface impurity concentration and their type.  Furthermore, the sensitivity of  interface spin-Hall to the interface potential $U_0$ provides a means of controlling spin-torque with gating.

In conclusion, our study demonstrates that the impurities at the metal-insulator interface significantly increase the variety of surface spin effects. The skew scattering from these impurities induces spin accumulation that extends up to the mean free path. When combined with spin relaxation, it gives rise to the interface spin Hall effect, which converts charge currents to spin currents at the metal-insulator boundary.

\section{Acknowledgements}
We have received funding from the European Community's
H2020 Programme under Grant Agreement INTERFAST
(H2020-FET-OPEN-965046), and from the Slovenian Research
Agency Program No.P1-0040.

\appendix

\section{spin polarization near the clean interface}
\label{App1}

Our starting point is the  Hamiltonian described by Eq.~(\ref{Ham}).
We are interested in its solutions when electron energy $\varepsilon$ is below the barrier $U_0$.
We start with  rotation in the spin space with the matrix:
\begin{equation}
U_\sigma=
\begin{pmatrix}
\tilde{k}_{\perp} & -\tilde{k}_{\perp}\\
1 & 1
\end{pmatrix}
\label{rotation}
\end{equation}
here $\tilde{k}_{\perp}=(ik_x+k_y)/k_{\perp}$, $k_{\perp}=\sqrt{k_x^2+k_y^2}$.
After canonical transformation $U^{-1}_\sigma HU_\sigma$ the Hamiltonian becomes diagonal:
\begin{equation}
H=-{\hbar^2\over{2m}}{d^2\over{dz^2}}+{\hbar^2k_{\perp}^2\over{2m}}+U_0 \theta(z) \mp \gamma U_0\delta(z)k_{\perp}
\label{hamil}
\end{equation}

\begin{widetext}
Using Eq.(\ref{hamil}) we find the eigenfunctions:
\begin{equation}\label{Psi}
\psi_{\pm}(\mathbf{r}_{\perp},z)  = \begin{cases}
A_{\pm}\begin{pmatrix}
\pm\tilde{k}_{\perp}\\
1
\end{pmatrix} \sin{(k_{\pm}z-\phi_{\pm})}\exp{(i\mathbf{k}_{\perp}\mathbf{r}_{\perp})}, z<0 \\
-A_{\pm}\begin{pmatrix}
\pm\tilde{k}_{\perp}\\
1
\end{pmatrix} \sin{(\phi_{\pm})}\exp{(i\mathbf{k}_{\perp}\mathbf{r}_{\perp})}\exp{(-\kappa z)}, z>0
\end{cases}
\end{equation}
Here $A_{\pm}=\sqrt{{\kappa_{\pm}\over{\kappa_{\pm}L+1}}}$, $\kappa=\sqrt{{2m(U_0-\epsilon_{\pm})\over{\hbar^2}}}$, $\kappa_{\pm}=\kappa\mp{2m\gamma { U_0} k_{\perp}\over{\hbar^2}}$,  $\epsilon_{\pm}={\hbar^2k_{\pm}^2\over{2m}}$, $L$ is the size of the sample in z direction that corresponds to the thickness of heavy metal layer. $k_\pm$ is electron wavevector in $z$-direction that will become spin-dependent after effects of finite $L$ will be taken into account. The phase shift $\phi_{\pm}$ is determined by the equation $\tan{\phi_{\pm}}=k_{\pm}/\kappa_{\pm}$. The energy is
\begin{equation} \label{Epm}
\varepsilon_{\pm}={\hbar^2k_{\pm}^2\over{2m}}+{\hbar^2k_{\perp}^2\over{2m}}
\end{equation}

\end{widetext}

To describe the boundary conditions for Boltzmann equation it is enough to consider the infinite sample $L \rightarrow \infty$.  In this case $k_+ = k_- = k_z$,
$\varepsilon_+ = \varepsilon_- = \varepsilon_k$
and the solution of Schrodinger equation exists for an arbitrary incident wave.
\begin{equation} \label{PsiInc}
\begin{pmatrix}
a\\
b
\end{pmatrix}\exp{(ik_z z)}\exp{(i\mathbf{k}_{\perp}\mathbf{r}_{\perp})}
\end{equation}
Here $(a,b)$ is an arbitrary spinor. According to Eq.~(\ref{Psi}) its reflected wave is described as follows
\begin{equation}
\hat{r}(k)\begin{pmatrix}
a\\
b
\end{pmatrix}\exp{(-ik_z z)}\exp{(i\mathbf{k}_{\perp}\mathbf{r}_{\perp})}
\end{equation}
Here
\begin{multline}
\hat{r}(k) = \exp{(i\pi+2i\phi_0)} \times \\  (\cos{(\Delta\phi)}\hat{\sigma}_0+i\sin{(\Delta\phi)}\hat{\sigma}_k)
\end{multline}
\begin{equation}
\hat{\sigma}_k={k_y\over{|k_{\perp}|}}\hat{\sigma}_x-{k_x\over{|k_{\perp}|}}\hat{\sigma}_y
\end{equation}

Interestingly, to account for Rashba-Edelstein effect at the clean interface, it is important to consider $L$ to be finite.
Here we assume the boundary conditions at $z=-L$:  $d\psi(\mathbf{r}_{\perp},z)/dz|_{z=-L}=0$. It leads the following equations for $k_\pm$ and  $\phi_{\pm}$:
\begin{multline} \label{kpm}
k_{\pm}={\pi\over{2L}}(2n+1)-{\phi_{\pm}\over{L}}, \\
\phi_{\pm}=\arctan{ \left[{(n+1/2)\pi-\phi_{\pm}\over{L\kappa_{\pm}}}\right]},
\end{multline}
where $n$ is arbitrary integer. Eqs.~(\ref{Epm},\ref{kpm}) show that when $L$ is finite the discrete levels for spin up and down are different, that will later lead to
the current-induced spin polarization.

To account for in-plane electric current we introduce the distribution function of the electrons $f({\bf k}) = f_0(\varepsilon-v \hbar k_x)$.
Here $f_0$  is the Fermi function, $v$ is the drift velocity that is assumed to be along $x$-direction.
The current density $j_x(z)$ is defined as follows:
\begin{multline} \label{jx1}
j_x(z) = \frac{e}{ L}\sum_n\int{d\mathbf{k}_{\perp}\over{(2\pi)^2}}{ 2 \hbar  k_x\over{m}} \times \\ \bigl(\sin{(k_+z-\phi_+)}^2f_0(\varepsilon_+-\hbar vk_x)
\\ +\sin{(k_-z-\phi_-)}^2f_0(\varepsilon_- - \hbar vk_x)\bigr)
\end{multline}
Note, that although the distribution $f({\bf k})$ does not depend on spin, the energies $\varepsilon_{\pm}$ do.

Eq.~(\ref{jx1}) can be simplified by expanding Fermi functions over small $vk_x$ and integrating over the angle of ${\bf k}_\perp$.
\begin{multline} \label{jx2}
j_x(z) = {ev {\hbar^2} \over{ 8   \pi mTL}}\sum_n\int dk_{\perp}k_{\perp}^3  \times \\ \left[\frac{\sin^2(k_+z-\phi_+)}{{\cosh^2{({\varepsilon_+-\mu\over{2T}})}}}
+\frac{\sin^2(k_-z-\phi_-) }{{\cosh^2{({\varepsilon_--\mu\over{2T}})}}} \right]
\end{multline}
Here $T$ is temperature and $\mu$ is the chemical potential. Eq.~(\ref{jx2}) describes the distribution of current density near the interface.

To describe the spin polarization we take into account that
\begin{equation}
(\pm \tilde{k}^*_{\perp},1)\sigma_y\begin{pmatrix}
\pm \tilde{k}_{\perp}\\
1
\end{pmatrix}
=\mp 2k_x/k_{\perp}
\end{equation}
It allows us to derive the expression for the distribution of spin polarization $m_y(z)$
\begin{multline}
m_y(z) = {\frac{1}{L}} \sum_n\int{d\mathbf{k}_{\perp}\over{(2\pi)^2}}{2k_x\over{k_{\perp}}} \times \\ \Bigl[\sin{(k_+z-\phi_+)}^2f_0(\varepsilon_+ - \hbar vk_x)
\\ -\sin{(k_-z-\phi_-)}^2f_0(\varepsilon_--\hbar vk_x) \Bigr]
\end{multline}
After expanding it over small $vk_x$ and performing the angle integration we obtain:
\begin{multline} \label{sigy}
m_y(z) =  {v\hbar \over{8 \pi T L }}\sum_n\int dk_{\perp}k_{\perp}^2  \times \\ \left[
\frac{\sin^2{(k_+z-\phi_+)}}{{\cosh^2{({\varepsilon_+-\mu\over{2T}})}}}
 -
 \frac{\sin^2{(k_-z-\phi_-)}}
 {{\cosh^2{({\varepsilon_--\mu\over{2T}})}}} \right]
\end{multline}
This formula describes the spin accumulation near the interface when current is flowing parallel to the surface. Fig.~\ref{fig:app} represents calculated current density and the accumulated spin density
as a function of $z$ (Rashba-Edelshtein effect). The current density is normalized to $j_{x,0} = evk_F^3/3\pi^2$ which is the current density far from interface.
Fig.~\ref{fig:app} shows that in a clean sample the spin polarization exists only at the distances $\sim \hbar/p_F$ from the surface where the current is modified by quantum effects.

 Note also, that the wavefunctions (\ref{Psi}) do not include density flux ${\rm Im}\psi_{\pm}^* \nabla \psi_\pm = 0$ and the spin density (\ref{sigy}) is not accompanied by spin current.

\section{spin-dependent scattering at the interface impurities}
\label{App2}

In the general case, impurity scattering requires density matrix formalism for its description. The usual procedure for such a description
starts with the density matrix $\hat{\rho}({\bf k})$ diagonal in the space of wavefunctions $\Psi_{\bf k}$ and considers non-diagonal terms $\hat{\rho}({\bf k},{\bf k}')$ as a small
perturbation. However, the terms $\hat{\rho}({\bf k})$ diagonal in the space of $\Psi_{\bf k}$ are still $2\times 2$ matrices in spin space.

The master equation for the density matrix reads
\begin{equation}\label{A1}
\frac{\partial \hat{\rho}}{\partial t} = \frac{i}{\hbar} (\hat{\rho} \hat{V} - \hat{V}\hat{\rho}).
\end{equation}
Here $V$ is the impurity energy that includes both potential energy and spin-orbital correction (\ref{Vimp-all}).

\begin{widetext}

\begin{figure}[b]
\includegraphics[angle=0,width=0.8\linewidth]{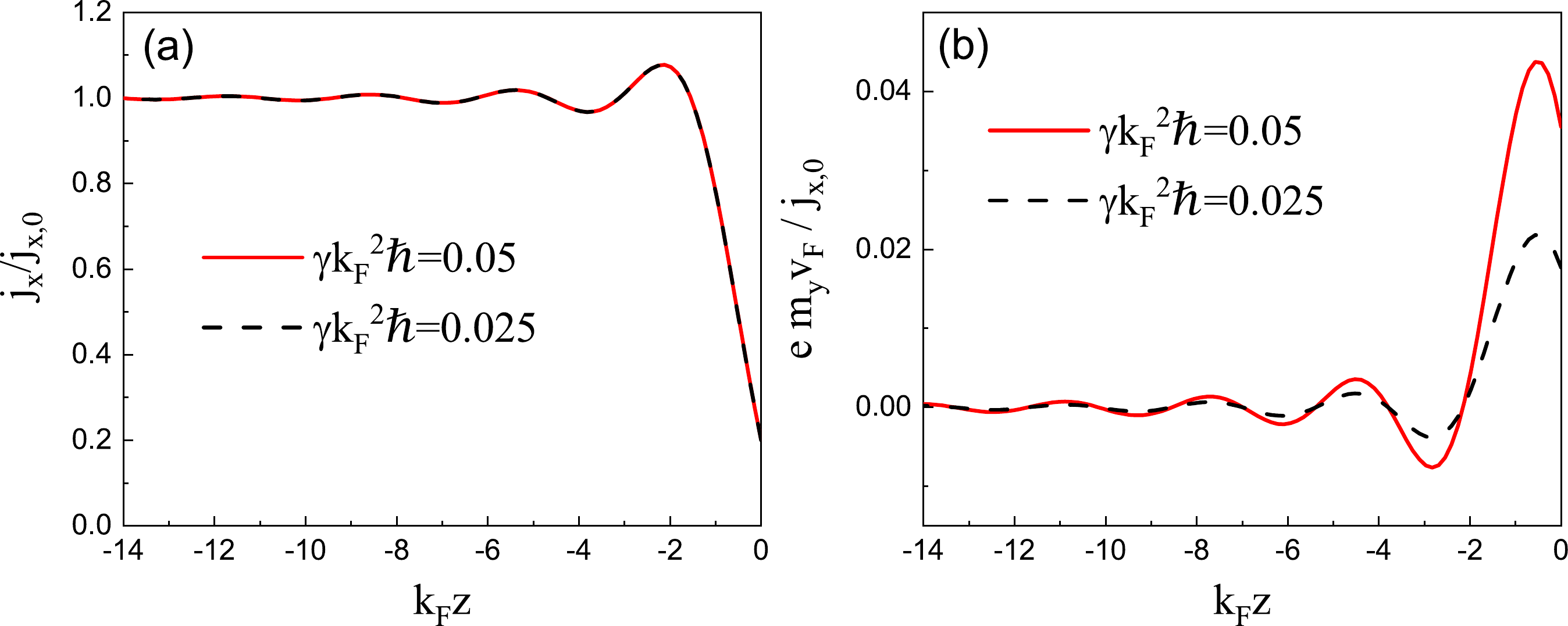}
\caption{The distribution of the current density near the surface (a) and spin density as a function of z (b).}
\label{fig:app}
\end{figure}

We assume the slow variance of diagonal matrix elements compared to the oscillation frequencies $(\varepsilon_{\bf k} - \varepsilon_{{\bf k}'})/\hbar$.
It allows us to derive expression for the perturbation $\hat{\rho}({\bf k},{\bf k}')$ from Eq.~(\ref{A1})
\begin{equation}\label{A2}
\hat{\rho}({\bf k},{\bf k}') = \frac{\exp\left(i \frac{\varepsilon_{\bf k} - \varepsilon_{\bf k'}}{\hbar} t\right)}{\varepsilon_{\bf k} - \varepsilon_{\bf k'} - i\delta}
\times
 \left( \hat{\rho}({\bf k})\hat{V}({\bf k}{\bf k}')  - \hat{V}({\bf k}{\bf k}') \hat{\rho}({\bf k}') \right)
\end{equation}
This equation should be substituted back to Eq.~(\ref{A1}) where we now keep only the non-oscillating terms
\begin{multline}\label{A3}
\frac{\partial \hat{\rho}({\bf k})}{\partial t}  = \sum_{\bf k'} \frac{i/\hbar}{\varepsilon_{{\bf k}'} - \varepsilon_{\bf k} - i\delta} \times
\left( \hat{V}({\bf k}{\bf k}') \hat{V}({\bf k}'{\bf k})\hat{\rho}({\bf k})  - \hat{V}({\bf k}{\bf k}') \hat{\rho}({\bf k}') \hat{V}({\bf k}'{\bf k})   \right) - \\
\frac{i/\hbar}{\varepsilon_{{\bf k}} - \varepsilon_{{\bf k}'} - i\delta} \times
\left( \hat{V}({\bf k}{\bf k}') \hat{\rho}({\bf k}') \hat{V}({\bf k}'{\bf k}) - \hat{\rho}({\bf k}) \hat{V}({\bf k}{\bf k}') \hat{V}({\bf k}'{\bf k})  \right)
\end{multline}
Here the real part of $1/(\varepsilon_{{\bf k}'} - \varepsilon_{\bf k} - i\delta)$ corresponds to the small modification of electron states due to impurity potential that can be neglected.
Imaginary part leads to actual transitions between states. They can be described with the equations
\begin{equation}\label{rho}
\frac{\partial\rho_{ij}({\bf k})}{\partial t} = \int \frac{V d{\bf k'}}{(2\pi)^3}\left(
W_{ij,lm}^{(in)} ({\bf k},{\bf k'}) \rho_{lm}({\bf k}') \right.
\left. - W_{ij,lm}^{(out)} ({\bf k},{\bf k'})\rho_{lm}({\bf k})
\right)
\end{equation}
\begin{equation}\label{win}
{W}_{ij,lm}^{(in)} ({\bf k},{\bf k'}) \rho_{lm}({\bf k}') = \frac{2\pi}{\hbar}N_I
\times V_{il}({\bf k},{\bf k}') \rho_{lm}({\bf k}') V_{mj}({\bf k}',{\bf k}) \delta(\varepsilon_k - \varepsilon_{k'})
\end{equation}
\begin{equation}\label{wout}
{W}_{ij,lm}^{(out)} ({\bf k},{\bf k'}) \rho_{lm}({\bf k}) = \frac{\pi}{\hbar}N_I
\times
\Bigl[
V_{in}({\bf k},{\bf k}')  V_{nl}({\bf k}',{\bf k}) \rho_{lm}({\bf k}) \delta_{jm} +
 \rho_{lm}({\bf k}) V_{mn}({\bf k},{\bf k}')  V_{nj}({\bf k}',{\bf k}) \delta_{il}
\Bigr]
\delta(\varepsilon_k - \varepsilon_{k'})
\end{equation}
Here $N_I$ is the total number of impurities.
$W_{ij,lm}^{(in)} ({\bf k},{\bf k'})$  describes the electrons scattering from the state ${\bf k}'$ to the state ${\bf k}$. Because both the initial and the final states of electrons are described with $2\times2$ density matrix, $W_{ij,lm}^{(in)} ({\bf k},{\bf k'})$ is, therefore,
 a four-dimensional $2\times2\times2\times2$ matrix.  ${W}_{ij,lm}^{(out)} ({\bf k},{\bf k'})$  is the out-scattering term that
stands for the backward transition from ${\bf k}$ to ${\bf k}'$. It is different from ${W}_{ij,lm}^{(in)}({\bf k},{\bf k'})$  because the spin is not conserved during the scattering.

For a given density matrix $\hat{\rho}({\bf k}')$ of incident electrons,
\begin{equation}\label{Wsc}
{W}_{ij,lm}^{(in)} ({\bf k},{\bf k'}) \rho_{lm}({\bf k}')
\end{equation}
represents the properties of the scattered ones.
In particular the results shown in  Fig.~\ref{fig:scat}(b,c) are calculated with Eq.~(\ref{Wsc}) where $\hat{\rho}({\bf k}') = (\hat{1} \pm \sigma_y)/2$ and ${\bf k}' = (p_F/\sqrt{2}\hbar)(1,0,1)$. The scattering probabilities correspond to ${\rm Tr} \widehat{W}^{(in)} ({\bf k},{\bf k'}) \hat{\rho}({\bf k}')  =  {W}_{ii,lm}^{(in)} ({\bf k},{\bf k'}) \rho_{lm}({\bf k}')$ and the spin polarization vector ${\bf s}$ to
${\rm Tr} \bm{\sigma}  r({\bf k}) \widehat{W}^{(in)} ({\bf k},{\bf k'}) \hat{\rho}({\bf k}') r^+({\bf k})     =  r^+_{ii_1}({\bf k}) \bm{\sigma}_{i_1i_2} r_{i_2j}({\bf k}) {W}_{ji,lm}^{(in)} ({\bf k},{\bf k'}) \rho_{lm}({\bf k}')$.

\section{Boundary conditions for the Boltzmann equation}
\label{App3}
The approach (\ref{rho}) is based on the wavefunctions $\Psi_\alpha({\bf k})$ defined in the main text. They correspond to the coherence of incident and specularly reflected electrons. However, in the bulk of the film this coherence is lost due to the scattering at the bulk impurities and The Boltzmann equation approach is based on the distribution function $\hat{f}({\bf p})$ that neglects such a coherence.

To relate the two approaches we consider the spin polarization generated per unit time. Because the kinetic equation is linear we can decouple it into $G_{\alpha}({\bf p}) d{\bf p}$ - the spin polarization in the $\alpha$-direction  related to the reflected electrons with the momentum ${\bf p}$. In terms of Boltzmann equation it is expressed as follows
\begin{equation}\label{Gf}
G_\alpha({\bf p}) d{\bf p} = \frac{d {\bf p}}{(2 \pi \hbar)^3} {\rm Tr} \hat{\sigma}_\alpha \hat{f}_2({\bf p}) \frac{|p_z|}{m}S
\end{equation}
Here $\hat{f}_2({\bf p})$ is taken at the interface and is assumed not to depend on the exact point of the interface. $S$ is the interface area. Note that $p_z$  is negative for the reflected electrons.

In terms of Eq.~(\ref{rho}) $G_{\alpha}({\bf p}) d{\bf p}$ is equal to
\begin{equation}\label{Grho}
G_\alpha({\bf k}) d{\bf k} = \frac{V d{\bf k}}{(2\pi)^3}  {\rm Tr} \hat{r}_{\bf k}^+ \hat{\sigma}_\alpha  \hat{r}_{\bf k}  \times
\int \frac{V d{\bf k}'}{(2 \pi)^3} \left(  W_{ij,lm}^{(in)}({\bf k},{\bf k}') \rho_{lm}({\bf k}')  - \right.
 \left. W_{ij,lm}^{(out)} ({\bf k},{\bf k}') \rho_{lm}({\bf k}) \right)
\end{equation}

When $\hat{f}_2 \ll \hat{f}_1\ll \hat{f}_0$ one can substitute $\hat{\rho}({\bf k})$ with $\hat{f}_0(\hbar k_x,\hbar k_y, \hbar |k_z|) + \hat{f}_1(\hbar k_x,\hbar k_y, \hbar |k_z|)$ in the r.h.s. of Eq.~(\ref{Grho}). However,
$\hat{f}_0$ does not lead to spin polarization and can be dropped.

It allows us to derive the following boundary condition for $\hat{f}_2$ at $z=0$.
\begin{equation}\label{f2}
\hat{f}_2({\bf p}) = \frac{m V}{S|p_z|}  \hat{r}({\bf p}) \left( \int\frac{V d{\bf p}'}{(2\pi \hbar)^3}  \times  \right.
\left.
\widehat{\cal W} \left(\frac{{\bf p}}{\hbar},\frac{{\bf p}'}{\hbar}\right)
\bigl(f_1({\bf p}') - f_1({\bf p}) \bigr)  \right) \hat{r}({\bf p})^+
\end{equation}
Here
\begin{equation}\label{calW}
{\cal W}_{ij} = W_{ij,ll}^{(in)} =  W_{ij,ll}^{(out)},
\end{equation}
and we took into account that $\hat{f}_1 = f_1 \hat{1}$.

\end{widetext}

\end{document}